\documentstyle[12pt,aasms4]{article}

\lefthead{Soifer et al}
\righthead{High Spatial Resolution Imaging of  Arp~220 }

\begin{document}

\title{
High Spatial Resolution Imaging of  Arp~220 from 3 - 25~$\mu$m$^{*}$
}

\author{B. T. Soifer$^{1}$, G.Neugebauer$^{1}$, K. Matthews$^{1}$, E. E. Becklin$^{2}$, M. Ressler$^{3}$,
\\  M. W. Werner$^{3}$,  A. J. Weinberger$^{1, 2}$ and E. Egami$^{1}$
}
\affil{$^{1}$Palomar Observatory, California Institute of 
Technology, 320-47,
Pasadena, CA 91125,\\
$^{2}$Department of Physics and Astronomy, University of California Los Angeles, 156205 Los Angeles, CA 90095,\\
$^{3}$Jet Propulsion Lab, 169-506, 4800 Oak Grove Dr., 
Pasadena, CA 91109,\\
Electronic mail: bts@mop.caltech.edu,
gxn@caltech.edu, kym@caltech.edu, becklin@astro.ucla.edu, mww@ipac.caltech.edu, ressler@cougar.jpl.nasa.gov, 
alycia@astro.ucla.edu, egami@mop.caltech.edu}

$^{*}$ Based in part on observations obtained at the W. M. Keck Observatory 
which is operated jointly by the California Institute of Technology and the 
University of California

\begin{abstract}
Images of  Arp~220  from 3.45 to 24.5~$\mu$m with 0.5$''$ resolution
are presented which clearly separate the nucleus into at least two
components. The western component is about three times more luminous
than the eastern component, but the silicate absorption in the fainter,
eastern component is roughly 50\% greater than the absorption in the
western component. Each component is marginally resolved.  The two
components seen at 24.5~$\mu$m are identified with the two radio
components. The western source most likely coincides with the high
extinction disk previously suggested to exist in Arp~220, while the
eastern nucleus is identified with a faint highly reddened source seen
in HST 2.2~$\mu$m NICMOS images. The two nuclei together account for
essentially all of the measured 24.5~$\mu$m flux density. Two models
are presented, both of which fit the observations. In one, the majority
of the total luminosity is produced in an extended star formation
region and in the other most of the luminosity is produced in the
compact but extincted regions associated with the two
nuclei seen at 24.5~$\mu$m. In both pictures, substantial luminosity at
100~$\mu$m emerges from a component having a diameter of 2--3$''$
($\sim$1kpc).

\end{abstract}
 
\newpage

\section {Introduction}

Since its identification as a highly luminous galaxy in the infrared
(\cite{Soifer84})  Arp~220~=~IC~4553/4 has undergone extensive
observational studies with the  recognition that it is the closest of
the ultraluminous infrared galaxies (\cite{Sanders88}).  Its bolometric
luminosity of 1.2$\times10^{12}L_{\sun}$ is  in the range of quasar
luminosities, but most recently Arp~220 has been cited as a prototype
for the appearance of forming galaxies at high redshifts
(\cite{vanderwerf98}).

The most striking property of  Arp~220 is the very strong infrared
emission  that rises rapidly from $\sim$0.5~Jy at 12~$\mu$m to $\sim$10~Jy
at 25~$\mu$m to a peak flux density of $\sim$100~Jy  at
60~$\mu$m (\cite{Soifer84}; \cite{Soifer87}).  This emission is
generally attributed to thermal emission by dust heated by an
underlying luminosity source; the dust emission at these wavelengths
accounts for more than 98\% of the bolometric luminosity of the
source.  Wynn-Williams \& Becklin (1993) showed that the size of the
source at 32~$\mu$m was $<$2$''$,  and that the
source was centered on the optically obscured nucleus of the galaxy.
This requires the luminosity source to be either a dense starburst, a
heavily dust enshrouded AGN or some combination of these sources.

Recent observational studies of  Arp~220 have revealed new
details of the inner few arc-seconds or 1~kpc in this system. 
At 2.2~$\mu$m,
Graham et al. (1990)  resolved the nuclear region into two
nuclei separated by 0.95$''$, and suggested that the two nuclei were
the remnants of the merging galaxy nuclei that were inferred from the
large tidal tails seen in optical imagery (\cite{Arp66};
\cite{Toomre72}).  Larkin et al. (1995) established from near-infrared
spectroscopy of the nuclei that they appeared to be orbiting and placed
a lower limit of 1.5$\times10^9 M_\odot$ on the total mass of the
orbiting nuclei.  Finally, Scoville et al. (1998) have presented NICMOS
data from 1.1--2.2~$\mu$m that resolve the entire region into several
sources.

Condon et al. (1991) and Baan \& Haschick (1995) separated the nucleus
into two peaks in the radio continuum and Scoville et al. (1998)
matched the radio peaks with two peaks seen in the 2.2~$\mu$m
NICMOS images. Smith et al.  (1998) have reported VLBI observations of
Arp~220 at 18 cm with resolution of $\sim$5 mas that resolve the
continuum of the separate nuclei into discrete point sources that they
interpret as supernova remnants within the nuclear starbursts.  Downes
\& Solomon (1998) and Sakamoto et al. (1998) have reported new
millimeter wave interferometry with $\sim$0.5$''$ resolution that has
resolved gaseous disks associated with each of the nuclei.

Previous mid-infrared imaging of  Arp~220 has been reported by
Keto et al. (1992) and Miles et al. (1996).  Keto et al. reported
observations in a broad band from 8--13~$\mu$m that constrain
the nuclear region to be less than 1.3$''\times0.9''$.  Miles et al.
reported observations at 8.8 and 12.5~$\mu$m with resolutions of
1.0$''$ and 0.7$''$ that showed that the nuclear region at these
wavelengths is basically a double source elongated mostly
east-west with the western source slightly resolved and having
1.6 and 3.3 times the flux density of the eastern source at 8.8
and 12.5~$\mu$m respectively.

In this paper we report infrared imaging at wavelengths from 3.45 to
24.5~$\mu$m with angular resolution of 0.3$''$ to 0.6$''$.  These data
directly probe the distribution of thermally emitting dust in the
nuclear region and trace the luminosity in this source at wavelengths
approaching the peak of the output for this system. Throughout this
paper, we will assume a Hubble constant of 75~km~s$^{-1}$~Mpc$^{-1}$.
At a redshift of 5400 km~s$^{-1}$, Arp~220 is then at a distance of
72~Mpc, so 1$''~\approx$~350~pc.

\section {Observations and Data Reduction}

The observations reported here were obtained primarily with the MIRLIN
mid-infrared camera (\cite{Ressler94}) at the f/40 bent Cassegrain
visitor port of the Keck II Telescope of the W. M.  Keck Observatory.
The camera used a 128x128 Si:As array and was configured to have a
plate scale of 0.138$''$~pixel$^{-1}$ for a total field of view of
17$''\times17''$.  Observations were obtained in eight infrared bands
from 7.9 to 24.5~$\mu$m.  In the 10~$\mu$m atmospheric window,
observations were made with six different intermediate band-pass
filters in order to sample the spectral features of dust emission in
this window.  Two additional filters covered the ends of the 20~$\mu$m
atmospheric window and were selected to provide wavelength coverage
reaching the longest wavelengths easily observed by ground-based
telescopes short  of submillimeter wavelengths. The central wavelengths
and band-passes of the filters are listed in  Table~1.

The MIRLIN observations presented here were made on the night of UT 18
March 1998. At each wavelength the observing procedure was the same.  A
chopping secondary with a square wave chop, an amplitude of  6$''$ in
the north-south direction and a frequency of approximately 4 Hz was
employed for fast beam switching. The frames at each chop position were
coadded in hardware resulting in two images. After an interval of
approximately a minute, the telescope was nodded east-west by 6$''$ and
a second set of two images was obtained in order to cancel residuals in
the sky and subtract telescope emission. This procedure was repeated a
number of times, giving a total time on the source between two and ten
minutes depending on the strength of the source. A nearby star,
$\beta$~Her, was observed just before the nucleus of Arp~220 was
observed in order to determine the point spread function (PSF). As
with all objects measured throughout the night,
$\beta$~Her was observed sequentially at all the relevant wavelengths; 
i.e. observations of $\beta$~Her at all wavelengths were completed 
before the
nucleus of Arp~220 was observed. The full width at half maximum (FWHM)
of $\beta$~Her, as well as that of five other stars measured throughout
the night,  was $\lesssim$0.2$''$ greater than the diffraction limit at
the shorter observing wavelengths, but was diffraction limited,
0.62$''$, at 24.5~$\mu$m.  Because the measured image size of the PSF
stars varied significantly over the night, we do not feel confident in
attributing differences $\lesssim$0.2$''$ between a measured source
size and the PSF size to an intrinsic source size except possibly at
wavelengths of 24.5~$\mu$m.

The data were reduced by differencing the  two images obtained within
the chop pairs at each nod location, and then coadding the resulting
four positive images, with their positions appropriately adjusted to a
common location, to yield a positive image centered in a field
approximately 6$''\times6''$.

The entire night was photometric and the sensitivity was based on
measurements of $\alpha$~Tau made at the beginning of the evening. The
photometric measurements of $\alpha$~Tau, and of the five stars made at
different wavelengths throughout the night, are consistent to
$\lesssim10\%$ with previous ground-based measurements and the color
corrected IRAS point source catalog flux densities. The flux density
corresponding to 0.0 mag was taken to follow the prescription given in
the Explanatory Supplement to the IRAS Catalogs and Atlases (Joint IRAS
Science Team 1989), and is listed in Table~1.

In addition to observations at wavelengths $\lambda\geq$ 7.9~$\mu$m,
observations at high spatial resolution at 3.45~$\mu$m ($L$) were
obtained of the nucleus of  Arp~220 using the Cassegrain infrared
camera on the 200-inch Hale Telescope of Palomar Observatory on 20 May
1997 in non-photometric conditions. Thirteen pairs of observations each
with 70 sec of integration (1000 coadded 0.07 sec frames) on the galaxy
and on nearby sky were obtained with a 64$\times$64 pixel sub-array of
the InSb array camera.  The pixel scale of the camera was
0.125$''$~pixel$^{-1}$ and hence the field of view was 8$''\times8''$.
Each image of the galaxy was sky subtracted and flat fielded.  Residual
sky levels were estimated for each image and subtracted.  Bad pixels
were corrected by interpolation.  The thirteen resulting images were
registered using the recorded positions of the offset guider and
averaged. In order to photometrically calibrate the final image,
further 3.45~$\mu$m images were obtained of the galaxy and of a
photometric standard on 5 June 1998, a photometrically good night, and
processed in the same manner as described above.

\section {Results}

The nuclear region of  Arp~220 was detected at all the wavelengths
where it was observed. The photometry of the nuclear region within a
4$''$ diameter beam and within  smaller beams (described below) within
the nuclear region is presented in Table~2.   The total detected flux
density of 16~mJy at 3.45~$\mu$m is consistent with the flux density of
22~mJy in a 5$''$ diameter beam at 3.7~$\mu$m as reported by Sanders et
al. (1988). At 24.5~$\mu$m, the flux density reported here of 9.8~Jy in
a 4$''$ diameter beam is likewise consistent with the flux density of
8.4~Jy measured by Wynn-Williams \& Becklin (1993) at 25~$\mu$m with a
5.7$''$ diameter beam and with the color corrected flux density of
10~Jy obtained by IRAS at 25~$\mu$m in a rectangular beam $\sim1'
\times 5'$.
 
A montage of images at all observed wavelengths is presented in
Figure~1a. As a comparison, contour maps of images of the star
$\alpha$~Tau, taken before the images of Arp~220 were obtained, are
shown in Figure~1b. In Figure~1a, the centroid of the western nucleus
has been set in the same position in each frame on the assumption that
the centroid of this source is spatially coincident at all
wavelengths.  As noted by Scoville et al. (1998) this is a dubious
assumption at the shorter wavelengths due to the presence of heavy
extinction in the nuclear region.  To facilitate a comparison of the
morphologies at all wavelengths, the location of the peak of the
eastern component in the image obtained at 24.5~$\mu$m has been marked
by a plus sign at the same relative location on all the images.  Since
these observations were made using the chopping secondary and nodding,
the images are not sensitive to structure on spatial scales
$\gtrsim5''$.

At all wavelengths the western nucleus is  detected strongly. An
eastern nucleus is also present  except at 9.7 and 10.3~$\mu$m where
the signal to noise ratios are not adequate to confidently locate it.
While the apparent peak of the eastern emission is at the same location
at 7.9, 8.8 and 17.9~$\mu$m, the peak brightness lies
 to the northeast of the 7.9~$\mu$m centroid at 3.45 and
11.7~$\mu$m, while it appears to be to the south by about 0.2$''$ at
12.5 and 24.5~$\mu$m. At 3.45~$\mu$m, the eastern nucleus seems to be
at the same location as the dominant eastern source seen by Scoville et
al. (1998) at 1.1 to 2.2~$\mu$m, while at the longer wavelengths, the
dominant eastern component apparently agrees in location more with a
faint source seen at 2.2~$\mu$m to the southeast by Scoville et al.
and with the center of the radio continuum sources seen by Smith et al.
(1998). The 3.45, 7.9 and 8.8~$\mu$m images show
that the eastern peak is itself elongated at a position angle of
$\sim$15 $\deg$; this is quite similar to the structure revealed in the
NICMOS images (\cite{Scoville98}).  This structure might also be
present at 11.7 and 12.5~$\mu$m, but is not clearly evident at longer
wavelengths.  At 24.5~$\mu$m, the two nuclei
 are separated by 0.94$ \pm 0.05''$ at a position  angle of 102 $\pm 5~
\deg$ in excellent agreement with the  separation of the mean of the
radio peaks given by Smith et al. of 0.94 $ \pm 0.01''$ at a
position angle of 102$ \pm 3 \deg $.

The complicated structure of the nuclear region makes accurate
photometry of its constituents impossible. In addition to the 4$''$
diameter circular beams, photometry was obtained on two smaller areas
within the nuclear region --- a 0.8$''$ diameter circular beam centered
on the western nucleus and a rectangular beam 0.7$''\times1.8''$
located as outlined in Figure~1a.  For the smaller beams, whose
dimensions are in some cases comparable to the size of the sources, the
sensitivity was adjusted by applying a similar beam to the star
observed just prior to the Arp~220 observations. An additional
correction of 33\% was applied to the 0.8$''$ diameter photometry of
the western source at 24.5~$\mu$m and, on the assumption the source is
the same size at 17.9~$\mu$m as at 24.5~$\mu$m, at 17.9~$\mu$m to
account for the fact that the source was extended while the PSF was
not. The accuracy of this correction depends on the observation that
the 24.5~$\mu$m observations throughout the night were diffraction
limited despite the variability of the seeing at shorter wavelengths.
The uncertainties given in Table~2 are dominated by photon statistics 
and do not include systematic uncertainties.   The
calibration of the magnitude of the standard star and the zero points
are probably uncertain at the 10\% level.  The uncertainties in the
small beam relative to each other and to the large beam measurements
are in the 10--20\% range due primarily to the uncertainties in the
detected source sizes.

The western source appears to be marginally resolved. The measured FWHM
of the western nucleus and the  FWHM of a star observed at each
wavelength just prior to the observations of Arp~220 are given in
Table~2 and are shown by the clear and hatched concentric circles
respectively in each panel of Figure~1a.  In all cases except
17.9~$\mu$m, the FWHM of the western nucleus exceeds that of the PSF
star indicating that the western source is resolved.  The 17.9~$\mu$m
measurement, where the source FWHM is smaller than that of the PSF,
apparently reflects a fluctuation in the seeing and the discrepancy can
perhaps be taken as an indication of the uncertainties in any size
determination presumably caused by seeing variations.

Figure~1a suggests that a low level of extended emission extends beyond
the two nuclei and has an overall extent of $\approx $2--3$''$
(700--1000~pc) at most wavelengths. The photometric measurements
indicate this emission is small. A comparison of the various
measurements given in Table~2 indicates that at most 10\% of the
emission at wavelengths longer than 3.45~$\mu$m is in this extended
component while a comparison between the emission at 24.5~$\mu$m and
the 25~$\mu$m IRAS measurements indicates that to within the
uncertainties of the calibration, the emission within the 1$' \times$
5$'$ IRAS beam is contained in the source area shown in Figure~1a.

Figure~2 shows the spectral energy distribution of the entire source
within the 4$''$ diameter beam as well as of the individual components
within the two smaller beams described above.  The spectral energy
distributions show deep minima near 10~$\mu$m which presumably indicate
strong absorption (Smith, Aitken \& Roche 1989). Other than the depth
of the absorption, the overall spectral energy distributions of the two
locations are quite similar within the uncertainties.

\section {Discussion}

The appearance of the nucleus of  Arp~220 in the mid-infrared is
complex because of the interplay of background light, embedded energy
sources and wavelength dependent emission and absorption by dust. This
complexity is already clearly apparent in the 1.1, 1.6 and 2.2~$\mu$m
NICMOS images of Scoville et al.  (1998) which show the effects of
large and highly variable extinction on the background starlight
emission from the nuclear environments. The mid-infrared data,  and the
24.5~$\mu$m data in particular, are, however, important in developing a
picture of Arp~220 because a significant fraction of the total
luminosity of the system is emerging at these wavelengths.  Further,
the strong absorption, presumably due to silicates (see e.g., Smith,
Aitken \& Roche 1989), that is apparent in the spectral energy
distributions of both nuclei implies that there is a significant column
of intervening cooler material in the line of sight to those regions.

The relative locations of the western and southeastern peaks at
24.5~$\mu$m are the same to within the uncertainties as  the relative
locations of the western clump of unresolved radio sources to the
southeastern clump (Smith et al. 1998).  Thus we associate the peaks
seen at 24.5~$\mu$m with the radio peaks seen by Condon et al. (1991),
Baan \& Haschick (1995) and Smith et al.  With this assumption,  the
southeastern peak is located within 0.1$''$ of the area associated with
the largest dust absorption as inferred by Scoville et al. (1998).
Based on the agreement between the separation and position angles of
the eastern and western nuclei at  24.5~$\mu$m and 1.3~mm  (Downes \&
Solomon 1998; Sakamoto et al. 1998) we also assume that the peaks at
these wavelengths are spatially coincident.  The agreement between the
distribution of 24.5~$\mu$m flux and the continuum at 1.3~mm is
important since the 1.3~mm emission is presumably optically thin so it
traces the total dust column density in the nuclear regions in this
system.
 
We attribute the strong dip near 10~$\mu$m (Figure 2) to absorption by
silicates. The eastern region has a significantly deeper absorption
than the western source around 10$\mu$m. The observations at
17.9~$\mu$m show a dip, a further signature of silicates, in both
regions, but especially in the eastern source. The ground-based
10~$\mu$m spectra of Smith, Aitken \& Roche (1989) as well as the ISO
spectra of Klaas et al.  (1997) show evidence for aromatic hydrocarbons
(PAH) in emission. The present data do not have adequate spectral
resolution to distinguish silicate absorption from an underlying
silicate/blackbody/PAH continuum, although the observations of Smith,
Aitken \&  Roche and Klaas, et al.  do suggest that PAH emission does
contribute to the continuum that is being absorbed.  The images where
the northeast elongation clearly appears were taken at wavelengths
where the very strong infrared emission features attributed to PAH
emission are known to appear; i.e., 3.4, 7.8, 8.7 and 11.5~$\mu$m in
Arp~220 [3.3, 7.7, 8.6 and 11.3~$\mu$m in the rest frame] (Willner et
al. 1977).  It is possible that the extended structure in Arp~220 is
traced most strongly in these features.

The interpretation of the 3.45--24.5~$\mu$m observations depends
critically on the size derived from the 24.5~$\mu$m observations.  The
measured size of the western nucleus at 24.5~$\mu$m, 0.73$''$, includes
a significant component from the diffraction limit of the telescope or
0.62$''$.  Subtracting 0.62$''$ in quadrature from the observed size
leads to a source size of 0.39$''$.  While the measured size of the PSF
was consistently 0.62$''$ during the observations reported here and we
believe the observations are nearly diffraction limited, there is a
possibility that seeing fluctuations could have increased the apparent
size of the source. The seeing contribution, as extrapolated from the
measured seeing at 12.5~$\mu$m, could be as large as 0.3$''$.   If
seeing contributed 0.3$''$ to the image diameter at 24.5~$\mu$m, the
source diameter is 0.25$''$.   Thus a range of source diameters at
24.5~$\mu$m of 0.25$''$ to 0.39$''$ is possible.

The 24.5~$\mu$m emission of the western source is approximately three
times that of the eastern source.  While the absorption is greater in
the eastern source, the overall spectral energy distributions of the
two nuclei do not suggest the eastern source is at a significantly
different temperature than the western one, and there is no strong
evidence that the eastern source is contributing  very differently to
the total luminosity of the system than inferred directly from the
observed 24.5~$\mu$m emission. For simplicity, in the following
discussion we will therefore assume that the western and eastern nuclei
have similar spectral energy distributions.

\subsection{Models}

The emission at wavelengths longer than 20~$\mu$m can be modeled in
several ways.  In the first scenario we  assume the largest source
diameter, 0.39$''$, and an optically thick nucleus that is unobscured at 
24.5~$\mu$m. This size is just large enough to contain all of the VLBI
radio point sources in the western nucleus (Smith et al. 1998),
suggestive that this size is appropriate for the compact western
nucleus.  If the measured 24.5~$\mu$m flux density in the western
source is due to an optically thick source of diameter 0.39$''$, the
24.5~$\mu$m brightness temperature is 85~K. An 85~K blackbody of
diameter 0.39$''$ emits 365~mJy at 1.3~mm. The dust continuum emission
from the western source measured at 1.3~mm by Sakamoto et al.  (1998)
is $\sim$140~mJy thus implying that the emissivity of the source is
0.5 at 1.3~mm, consistent with it being optically thick at
submillimeter wavelengths. If the eastern source has a similar spectral
energy distribution as the western source, the two nuclei together
account for essentially all of the flux density measured at 24.5~$\mu$m
in a 4$''$ diameter beam or by IRAS at 25~$\mu$m, but only 42~Jy or
40\% of the 104~Jy observed at 60~$\mu$m  by IRAS (Joint IRAS Science
Team 1989).  Similarly, an extrapolation to the far infrared implies
that the nuclei together produce $\sim$5$\times$10$^{11}L_{\sun}$
or $\sim$40\% of the bolometric luminosity of Arp~220.

In this model, the remaining continuum flux at 60--100~$\mu$m must come
from the surrounding environment. This surrounding region cannot
contribute significant flux at 24.5~$\mu$m  or 1.3~mm, must be
transparent at wavelengths shorter than 17.9~$\mu$m and yet must
contribute  $\sim$60\% of the bolometric luminosity of the system. This
would be an extended starburst region of prodigious luminosity ($\sim
7\times 10^{11} L_{\odot}$ ). We will assume that $\lesssim$10--15\% of
the emission at 24.5$\mu$m and 1.3~mm, i.e., flux densities
$\lesssim$1~Jy and $\lesssim$30~mJy respectively,  would go undetected
and could be in this component. We can fit the spectral energy
distribution around 60--100~$\mu$m with a source having a dust
temperature of 40~K and an emissivity $\epsilon\sim\nu^{2}$ where $\nu$
is the frequency of the emission.  To provide the excess flux of
$\sim$70 Jy at 100 $\mu$m the source must be larger than $\sim$2$''$ in
diameter. This size is also consistent with the source size measured
for Arp~220 at 350~$\mu$m by Benford (1998).  If the diameter of this
source is taken as 2.6$''$, or $\sim$900~pc, consistent with the
observed size of the extended CO emission in the source (Scoville, Yun
\& Bryant 1997), the optical depth at 100~$\mu$m is 0.7. The steep
emissivity law, $\epsilon\sim\nu^{2}$, is necessary to reduce the
predicted dust continuum at 1.3~mm to levels that would not be detected
in the existing maps (Sakamoto et al.  1998); the fit predicts a flux
density of 30~mJy for the 40~K component at 1.3~mm.  This emissivity
behavior is physically reasonable based on a lack of resonances at
wavelengths $>$~100~$\mu$m in the grains producing the submillimeter
continuum.  If the 40~K component is  distributed in a disk--like
structure, as will be discussed below, it can be nearly optically thick
at 100$\mu$m while allowing more compact structure to be seen at
shorter wavelengths.  The low emission of a blackbody of this low
temperature in the mid-infrared precludes using the present
observations to define this component better. The large extent and
luminosity of this component are also consistent with the distribution
of the CO emission, which places 70\% of the molecular gas outside the
compact nuclei (Sakamoto et al.) and with seeing both of the nuclei in
absorption at 10~$\mu$m.

The apparent diffuse emission at $\lambda < $ 17.9$\mu$m requires that
dust significantly warmer than 85~K surrounds the compact nuclei.  This
dust is almost certainly heated externally to the compact nuclei,
probably as a result of  a large extended starburst environment. This
provides a natural explanation of the low level extended mid-infrared
emission associated with this region.  It is reasonable that the
excitation of the PAH emission seen in Arp~220 (Smith, Aitken \& Roche,
1989; Klaas, et al. 1997;  Genzel et al.  1998) is associated with the
distributed starburst in the same way that similar features are seen in
other starburst galaxies such as M82 (e.g.  Willner et al 1977).  The
observed 7.9 and 8.8~$\mu$m flux densities in Arp~220 from Table~2
imply that the 8~$\mu$m luminosity $\approx$4$\times10^{10}L_{\odot}$.
Under the simple assumption that $\lambda L_{\lambda}$/$L_{bol}$ in
this extended starburst region is the same as that in M82 (Willner, et.
al., Helou et al 1998), we estimate a bolometric luminosity for this
region of 1--4$\times10^{11}L_{\odot}$ if the optical depth at 8~$\mu$m
is negligible.  An optical depth at 8~$\mu$m of $\sim$0.5--1.5 would
then be consistent with the observed flux. Such an 8~$\mu$m optical
depth is consistent with the 10~$\mu$m optical depth of $\tau\sim$7
for Arp~220 attributed to silicate absorption by Smith, Aitken \& Roche.

One possible problem with the extended starburst model is the faintness
of the Brackett-$\alpha$ line. DePoy, Becklin \& Geballe (1987)
observed a faint extended line with a 5$''$ diameter beam.  When they
corrected for a reddening of 50 mag of visual extinction, the strength
of the line corresponded to a starburst luminosity of about
2--3$\times$10$^{11}L_\odot$ (see Table~1 of DePoy et al.) which is 
substantially below the luminosity of $\sim
7\times 10^{11} L_{\odot}$ discussed above. Further, the
amount of extinction to an extended starburst region could be
comparable to A$_v$~=~50~mag if $\tau_{100\mu m}\sim$ 0.7.  An A$_v$ of
50~mag is the extinction inferred from the silicate absorption feature
corresponding to the total extinction in the central two sources.

The second scenario we present, which cannot be distinguished from the
first on the basis of these observations, follows from the assumption
that there is absorption at 24.5~$\mu$m, so the intrinsic emission of
the nuclei is attenuated by an opacity $\tau \approx$~1--2.  The
attenuating material is assumed to be in the proximity of the nuclei
and is heated and reradiates as well, so that the compact sources
provide the luminosity of Arp~220. In support of the assumption that
$\tau \gtrsim$1 at 24.5~$\mu$m is the fact that there appears to be
silicate absorption at 17.9~$\mu$m most clearly visible in the eastern
component (Figure~2).  In order for the compact nuclei to provide the
total luminosity of Arp~220, the temperature of the western source must
be between 102~K, if the source diameter is 0.39$''$, and 128~K, if the
diameter of the source is 0.25$''$. In the former case, the extinction
to the source is $\tau$~=~1.2 at 24.5 $\mu$m; in the latter case it is
$\tau$~=~1.5.  The eastern nucleus has been assumed to have the same
spectral energy distribution as the western nucleus and suffer similar
extinction.  In order to match the observations at 1.3~mm, the material 
surrounding the nuclei
must become optically thin at that wavelength; the calculated  values
of emissivity are very similar to those proposed for the previous
model. Since in this scenario the total bolometric luminosity is
provided by the compact nuclei, the need for an extended (starburst)
region is much diminished, but the long wavelength radiation comes from
the absorbing material in the neighborhood of the nucleus.

The overall energetics of the second model are shown in Figure~3; the
gross appearance is much the same as for the first scenario discussed.
The figure emphasizes the fact that, however the luminosity is 
generated, the 24.5~$\mu$m observations require  a component at a 
temperature 30--50~K and size measured in arc-seconds to radiate some 
significant fraction of the luminosity of Arp~220. Since both models 
are carefully
contrived to fit the present data, we cannot distinguish between the
two models on the basis of the available observations. The correct
picture is probably some combination of  the two pictures described 
above and involves some extinction at 24.5~$\mu$m and some starburst
activity extended on a 2--3$''$  scale providing some of the
luminosity of Arp~220.

\subsection{General Discussion}

Independently of the details of the models, the observations of compact
sources at 24.5~$\mu$m and 1.3~mm and the inference that even at
100~$\mu$m the sources must have substantial optical depth and
comparatively small size lead to the conclusion that we are seeing a
wide range of source temperatures and sizes with sources that are
optically thick at all observed wavelengths shorter than 25~$\mu$m.
The nuclear disks inferred from the millimeter CO observations of
Sakamoto et al (1998) and Downes and Solomon (1998) present a natural
way to accommodate these observations as well as the global feature
from Figure~1a that the images at all wavelengths from
3.45--24.5~$\mu$m show the same qualitative structure.

If the east and west sources are both disks viewed such that both are
reasonably close to the plane of the sky (i.e. significantly away from
edge on), the apparent global morphology will be the same at all
wavelengths, the size will increase with wavelength and at each
wavelength the source can appear to have significant optical depth.  
If
the disks are warped and thick, the outer disk could provide the
apparent cold silicate absorption through which the hotter material is
seen.  Thus we conclude that the infrared observations of Arp 220 are
entirely consistent with the tilted disks suggested from the millimeter
CO data.

The observed emission at wavelengths $<$17.9~$\mu$m is well in excess
of that predicted for either of the scenarios described above.
Specifically, at 12.5~$\mu$m an 85~K blackbody of 0.39$''$ diameter
emits 75~mJy, compared to the observed emission of 269~mJy, while if
the attenuation keeps increasing $\propto\nu^{2}$, the emission from
an attenuated 100--130~K blackbody is $<$1~mJy. Because dust opacity
increases to shorter wavelengths, this cannot be a result of seeing
deeper into a uniform distribution of material.  Undoubtedly, the
nuclear region is geometrically complex and must include non-uniform
extinction over both nuclei.  The disks seen by Sakamoto et al.
(1998) provide a natural means of allowing visibility into  regions
that appear optically thick at long wavelengths while allowing
radiation at shorter wavelengths to escape.

In any viable model, the luminosity of Arp~220, or a significant
fraction of it, is generated in a volume $\sim$10$^6$~pc$^3$, with an
energy output per unit volume $\gtrsim$100 times that in the extended
nuclear starburst in M82.  No direct observational evidence has been
presented for a non-stellar origin of the luminosity originating in the
nuclei of Arp~220.   High excitation infrared fine structure lines such
as [O IV] 24~$\mu$m and [Ne V] 14~$\mu$m have not been detected in
Arp~220 in recent observations with ISO SWS (Genzel et al. 1998).  The
present results show that the nuclei in Arp~220 are probably
optically thick even at 24.5~$\mu$m, and therefore the existence of a
central AGN in these nuclei cannot be probed even with these
mid-infrared lines.

\section {Summary and Conclusions}

High spatial resolution imaging of the nuclear regions of  Arp~220 from
3.45--24.5~$\mu$m has shown that the central source consists of at least
two compact sources with diffuse emission surrounding them.  Both 
compact sources are heavily obscured at 10~$\mu$m due to
silicate absorption.  The western nucleus is marginally spatially
resolved and provides, depending on the detailed  structure,  at least
30--70\% of the bolometric luminosity of Arp~220.  The source that
generates this luminosity must have a luminosity density $\gtrsim$100
times that of the starburst region in M82.  The optical depth
inferred for this source implies that there is likely no meaningful
probe of the power source short of radio wavelengths.  The large number
of supernova remnants reported by Smith et al. (1998) clearly
demonstrates that young stars must be contributing significantly to the
luminosity of this source. The eastern nucleus appears to contribute
about one-third the luminosity of the western nucleus, and appears to
be spatially extended in the northeast direction.

Despite the fact that there is significant extinction of
3.45--24.5~$\mu$m radiation in Arp~220, the presence of emission at
3.45--12.5~$\mu$m indicates that there are lines of sight to hotter
regions relatively free of extinction.  The two models we have
proposed, one in which the majority of the total luminosity is produced
in an extended star formation region and another in which the majority
is produced in two compact but extincted regions, both allow a geometry
in which short wavelength emission can escape.  Disks of material with
the range of temperatures described here which are inclined to our line
of sight provide a natural mechanism for achieving this.

If the extinction to the western source at 24.5~$\mu$m is
small, the western source  has a brightness temperature at 24.5~$\mu$m 
of 85~K; if the extinction is actually higher, it will be hotter.  
Both models
must therefore include hot (T~$>$85~K) embedded sources, and hence the
radiative transfer in both implies that the colder surrounding 
material,
which will emit most strongly at wavelengths greater than 60~$\mu$m, 
must
be extended.  If most of the luminosity is generated in the compact
sources, this extended emission will have a size $\gtrsim$1$''$ in 
radius
whereas if most of the luminosity is generated outside the nuclei, 
this
colder extended region must be at least 50\% larger.

Both of these models
 can reproduce the bolometric
luminosity of the galaxy if the flux is extrapolated to 60--100~$\mu$m
where the bulk of the luminosity of Arp~220 is observed.  Both are 
also consistent with the measured
sizes of the nuclear region in CO and continuum dust emission.

\section {Acknowledgments}

BTS, GN and KM are supported by grants from NASA and NSF. We thank J.
Aycock, R. Goodrich,  R. Moskitis and the entire Keck staff for their
help establishing the visitor port and obtaining these observations,
 and R. Chary, A. Evans, D.  Koerner, K.  Sakamoto, D.  Sanders and 
N.  Scoville
for helpful discussions about Arp~220 and these observations. The W.
M. Keck Observatory is operated as a scientific partnership between the
California Institute of Technology, the University of California and
the National Aeronautics and Space Administration. It was made possible
by the generous financial support of the W. M. Keck Foundation.  This
research has made use of the NASA/IPAC Extragalactic Database which is
operated by the Jet Propulsion Laboratory, Caltech under contract with
NASA.

\newpage
\figcaption{} (a) Contour maps are shown of  Arp~220 as observed at the
nine different wavelengths indicated. North is up and east is to the
left in all frames. The level of the lowest contour has been
arbitrarily set at the mode of the sky adjoining the galaxy plus three
times the rms noise of that sky. In order to emphasize the low level
features of the image, the lowest contour interval has been set to the
rms noise level of the sky; subsequent intervals increase by factors of
1.5.  At each wavelength the FWHM of the star $\beta$~Her taken just
before the observations of  Arp~220 is indicated in the hatched circle,
while the FWHM of the western nucleus is indicated by the clear circle.
The plus sign is located at the same place with respect to the western
nucleus in all the images; it is located at the peak of the eastern
nucleus at 24.5~$\mu$m. The rectangles  outline the area used for
photometry of the eastern source.

(b) Contour maps of the star $\alpha$~Tau taken before 
the Arp~220 observations are shown at the same scale as Figure~1a. 
The contours have been selected to correspond to those of the 
corresponding frames in Figure~1a when the extreme difference 
in brightness is taken into account. The bar over the wavelength 
identification corresponds to 1.44*wavelength/telescope diameter.

\figcaption{} The spectral energy distribution of a 4$''$ diameter beam
centered on the western nucleus is shown together with those of a 
0.8$''$ diameter beam centered on the western nucleus and a  
0.7$''\times1.8''$ rectangular beam at the position outlined in 
Figure~1a. The
flux densities have been corrected for the finite beam size as
described in the text.

\figcaption{} Various observations of Arp~220 are shown together with
the second model of obscured nuclei presented in the text.  The data
points include the present observations with a 4$''$ diameter beam
(MIRLIN), the   photometry from Klaas et al. (1997) (ISOPHOT), the
32~$\mu$m photometry from Wynn-Williams \& Becklin (1993) (IRTF),
sub-millimeter data from Rigopoulou, Lawrence \& Rowan-Robinson (1996)
(JCMT) and millimeter data from Sakamoto et al. (1998) (OVRO). For the
OVRO data, both the total and the western nuclear fluxes are plotted by
crosses. For the model, plotted as a solid line, we assume that almost
all the luminosity of Arp~220 comes from two compact nuclei, and that
the total luminosity of the galaxy is 1.3 times that of the western
nucleus, the ratio being determined from the 24.5~$\mu$m data.  A
blackbody temperature of 102~K was derived for the western nucleus with
the assumptions that the diameter of the nucleus is 0.39$''$, and that
$\nu f_{\nu}$ at the peak of this blackbody curve is 77\%  of $\nu
f_{\nu}$ at 60~$\mu$m calculated from the IRAS data.  A foreground
extinction was applied in the form $exp(-\tau_{24.5 \mu m} (\nu /
\nu_{24.5 \mu m})^2)$ with $\tau_{24.5 \mu m}=1.2$, which gives a good
fit to the western nucleus data points at 17.9 and 24.5~$\mu$m (not
plotted in the figure).  This extinction-applied curve was then
multiplied by 1.3 to take into account the contribution from the
eastern component, which was assumed to have the same spectral energy
distribution.  Finally, the result was multiplied by $(1-exp(-(\nu /1.5
THz)^2))$ to make the emitting region optically thin at the longer
wavelengths so that the curve matches the 1.3~mm OVRO data.  The result
is plotted as the dotted line.  To obtain the final fit (solid line),
emission from another lower temperature dust component in the form of
$B_{\nu}(T=40~K)(1-exp(-(\nu / 3 THz)^2))$  was added in (dashed line).
The size of the 40~K emission was assumed to be 4.6 $\Box ''$, which
was derived by equating the energy emitted by the 40~K  component with
the energy absorbed by the screen in front of the 102~K blackbody
nuclei.

\clearpage
\normalsize
\newpage
\begin{deluxetable}{ccc}
\tablewidth{0pt}
\tablecaption{Observing Bands}
\tablehead{
\colhead{Central}      &\colhead{Half Power}   &\colhead{Flux Density}\\
\colhead{Wavelength}   &\colhead{Width}        &\colhead{at 0.0 mag} \\
\colhead{$\mu$m}       &\colhead{$\mu$m}       &\colhead{Jy}   }

\startdata

  3.45      &0.24 &292 \nl
  7.9      &0.76 &63.3 \nl
  8.8      &0.87 &51.5 \nl
  9.7      &0.93 &42.7 \nl
 10.3      &1.01 &38.0\nl
 11.7      &1.11 &29.7\nl
 12.5      &1.16 &26.2\nl
 17.9      &2.00 &13.0\nl
 24.5      &0.76 &7.0\nl
\enddata
\end{deluxetable}

\begin{deluxetable}{cccccc}
\tablecolumns{6}
\tablecaption{Photometric Results}
\tablenum{2}
\tablehead{
\colhead{Central} & \multicolumn{3}{c}{Flux Density\tablenotemark{a}}    & \multicolumn{2}{c}{FWHM} \\
\colhead{Wavelength} & \colhead{Total \tablenotemark{b}} & \colhead{West \tablenotemark{c}}& \colhead{East \tablenotemark{d}}  & \colhead{West\tablenotemark{e}} & \colhead{PSF\tablenotemark{f}} \\
\colhead{$\mu$m} & \colhead{mJy}  & \colhead{mJy}  & \colhead{mJy}  &  \colhead{$''$} &  \colhead{$''$} }
\startdata
3.45 &	16 $\pm$ 2 &   5.0 $\pm$ 0.5 &	3.6 $\pm$ 0.3  &	0.69 &	0.42 \nl
7.9 &	521  $\pm$ 15 &	304 $\pm$ 3 & 	159 $\pm$ 4    &	0.45 &	0.38 \nl
8.8 &	155 $\pm$ 6 &	103 $\pm$ 1 &	58 $\pm$ 3     &	0.42 &	0.35 \nl
9.7 &	124 $\pm$ 14 &	91 $\pm$ 4 &	31 $\pm$ 4     &	0.52 &	0.38 \nl
10.3 &	59 $\pm$ 10 &	52 $\pm$ 2 &	12 $\pm$ 4      &	0.48 &	0.44 \nl
11.7 &	175 $\pm$ 8 &	121 $\pm$ 2 &	56 $\pm$ 2     &	0.42 &	0.39 \nl
12.5 &	404 $\pm$ 12 &	269 $\pm$ 3 &  102 $\pm$ 2     &	0.48 &	0.40 \nl
17.9 &	1170 $\pm$ 50 &	1030 $\pm$12 & 228 $\pm$ 14   &	0.52 &	0.56 \nl
24.5 &	9800 $\pm$ 200 & 7480 $\pm$ 100 & 2231 $\pm$ 45 &	0.73 &	0.62 \nl

\enddata
\tablenotetext{a}{The uncertainties are statistical; see the text.}
\tablenotetext{b}{Flux density in a 4$''$ diameter circular beam centered on the western source}
\tablenotetext{c}{Flux density in a 0.8$''$ diameter circular beam centered on the western source}
\tablenotetext{d}{Flux density in a 0.7$''~\times~1.8''$ rectangular beam shown in Figure~1a}
\tablenotetext{e}{FWHM of the western source}

\tablenotetext{f}{FWHM of the PSF,~ $\beta$~Her}
\end{deluxetable}

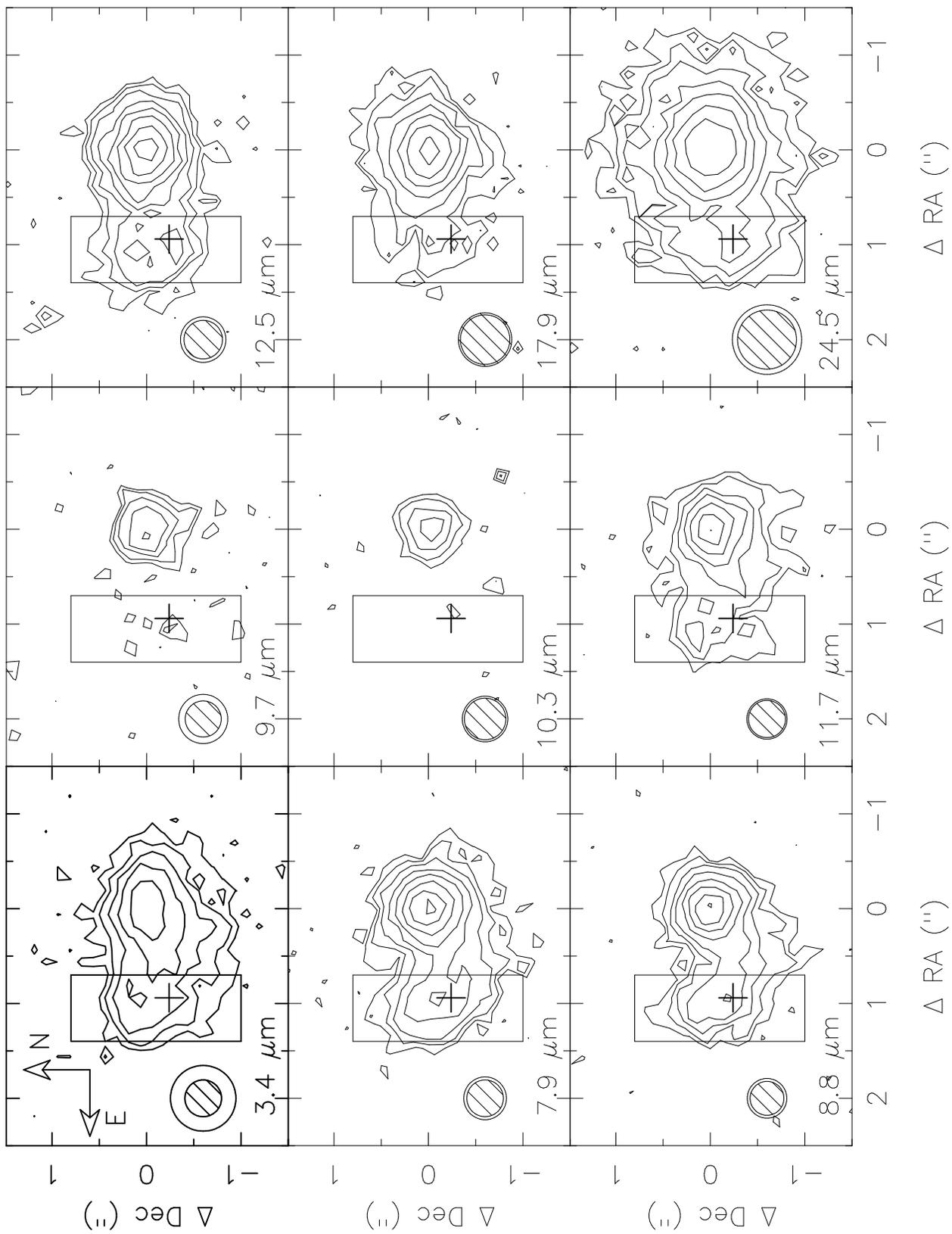
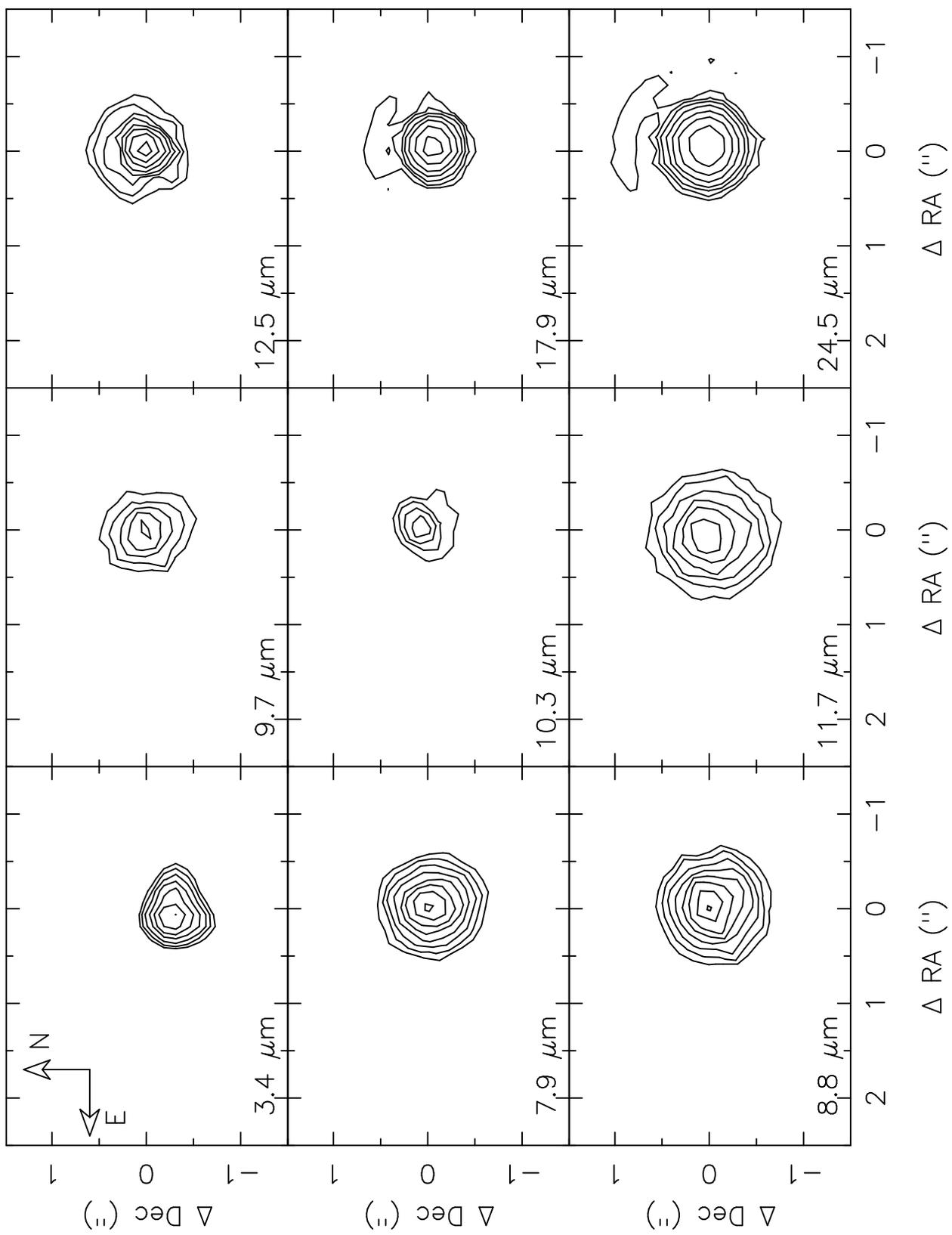
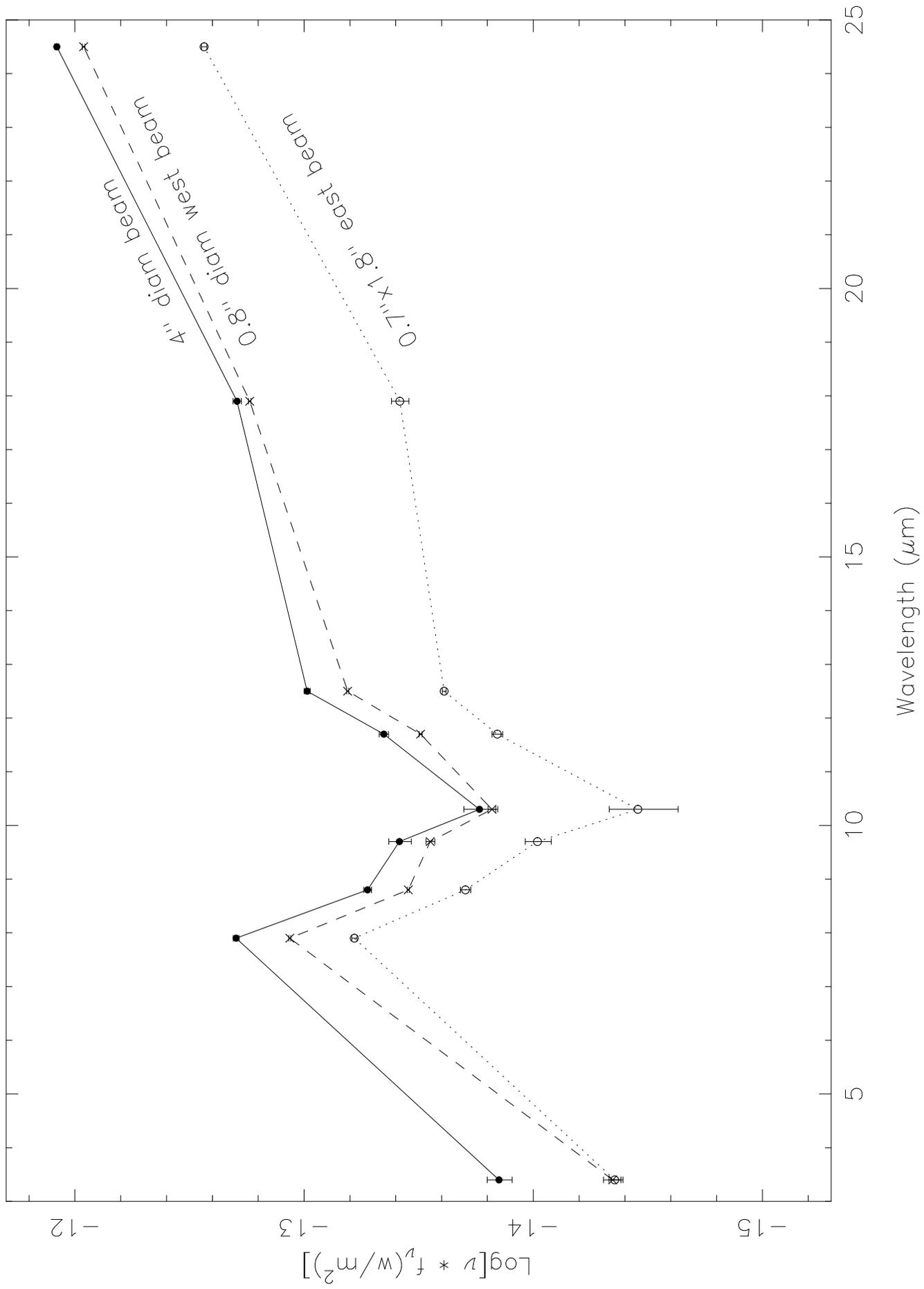
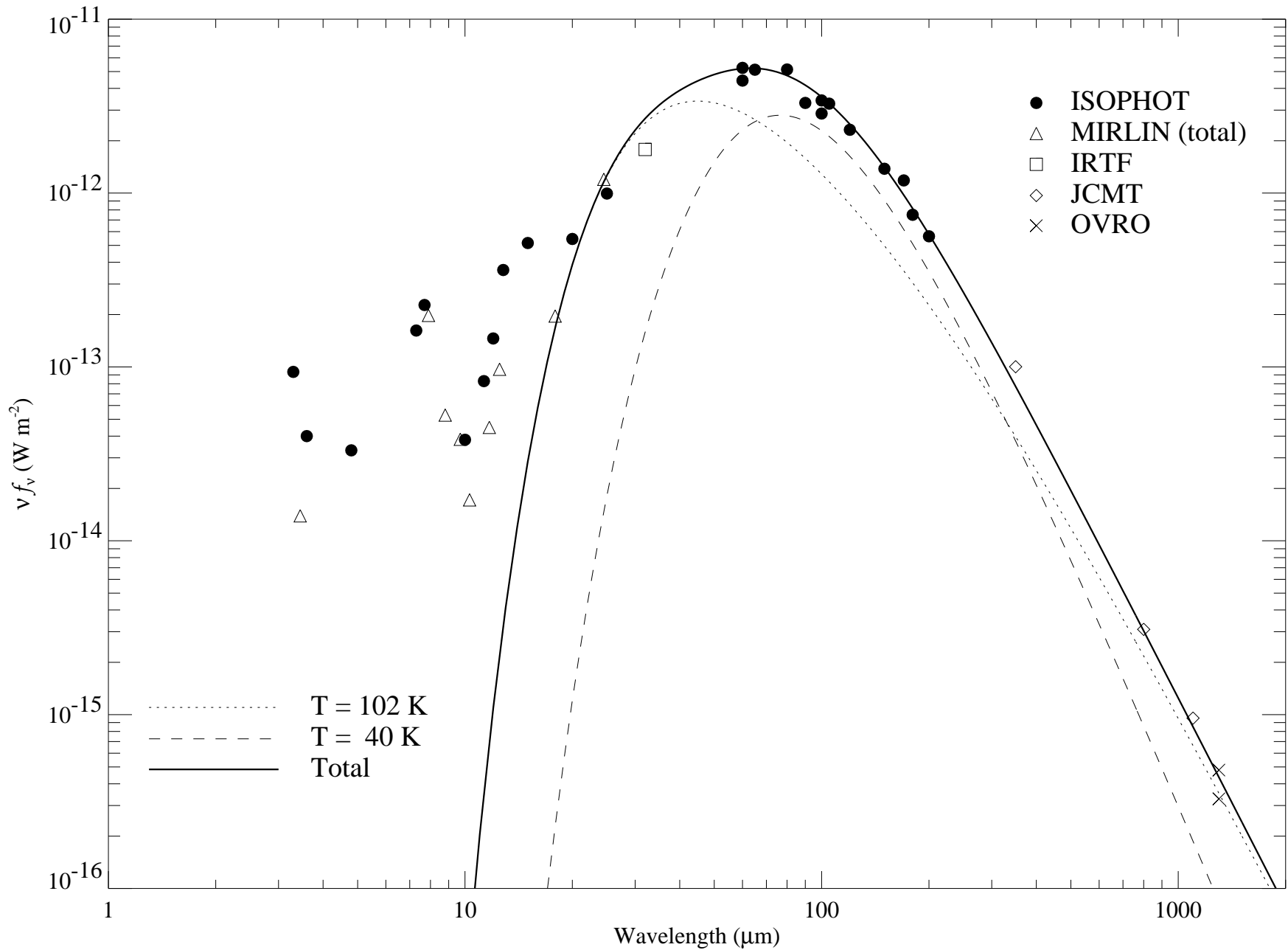
\end{document}